\documentclass{vgtc}                          

\graphicspath{{figures/}{pictures/}{images/}{./}} 

\usepackage{times}                     

\usepackage{tabu}                      
\usepackage{booktabs}                  
\usepackage{lipsum}                    
\usepackage{mwe}                       
\usepackage{tabularx}
\usepackage{booktabs}

\usepackage{mathptmx}                  

\onlineid{0}

\vgtccategory{Research}

\vgtcinsertpkg

\title{Exploring Organizational Readiness and Ecosystem Coordination for Industrial XR\thanks{\textcopyright\ 2026 IEEE. Personal use of this material is permitted. Permission from IEEE must be obtained for all other uses, in any current or future media, including reprinting/republishing this material for advertising or promotional purposes, creating new collective works, for resale or redistribution to servers or lists, or reuse of any copyrighted component of this work in other works. DOI: 10.1109/VRW70859.2026.00162}}

\author{Hasan Tarik Akbaba\thanks{e-mail: tarik.akbaba@tum.de}\\ 
        \scriptsize Technical University of Munich %
\and Efe Bozkir\thanks{e-mail: efe.bozkir@tum.de}\\ 
     \scriptsize Technical University of Munich
\and Anna Puhl\thanks{e-mail: anna.puhl@tum.de}\\   
     \scriptsize Technical University of Munich %
\and Süleyman Özdel\thanks{e-mail: ozdelsuleyman@tum.de}\\ 
     \scriptsize Munich Center for Machine Learning\\
     \scriptsize Technical University of Munich
\and Enkelejda Kasneci\thanks{e-mail: enkelejda.kasneci@tum.de}\\ 
     \scriptsize Munich Center for Machine Learning \\
     \scriptsize Technical University of Munich }

\abstract{
    Extended Reality (XR) offers transformative potential for industrial support, training, and maintenance; yet, widespread adoption lags despite demonstrated occupational value and hardware maturity. Organizations successfully implement XR in isolated pilots, yet struggle to scale these into sustained operational deployment, a phenomenon we characterize as the ``Pilot Trap.'' This study examines this phenomenon through a qualitative ecosystem analysis of 17 expert interviews across technology providers, solution integrators, and industrial adopters. We identify a ``Great Inversion'' in adoption barriers: critical constraints have shifted from technological maturity to organizational readiness (e.g., change management, key performance indicator alignment, and political resistance). While hardware ergonomics and usability remain relevant, our findings indicate that systemic misalignments between stakeholder incentives are the primary cause of friction preventing enterprise integration. We conclude that successful industrial XR adoption requires a shift from technology-centric piloting to a problem-first, organizational transformation approach, necessitating explicit ecosystem-level coordination.
} 


\begin{document}


\firstsection{Introduction}

\maketitle
\begin{figure}[!ht]
    \centering
    \includegraphics[width=0.9\columnwidth]{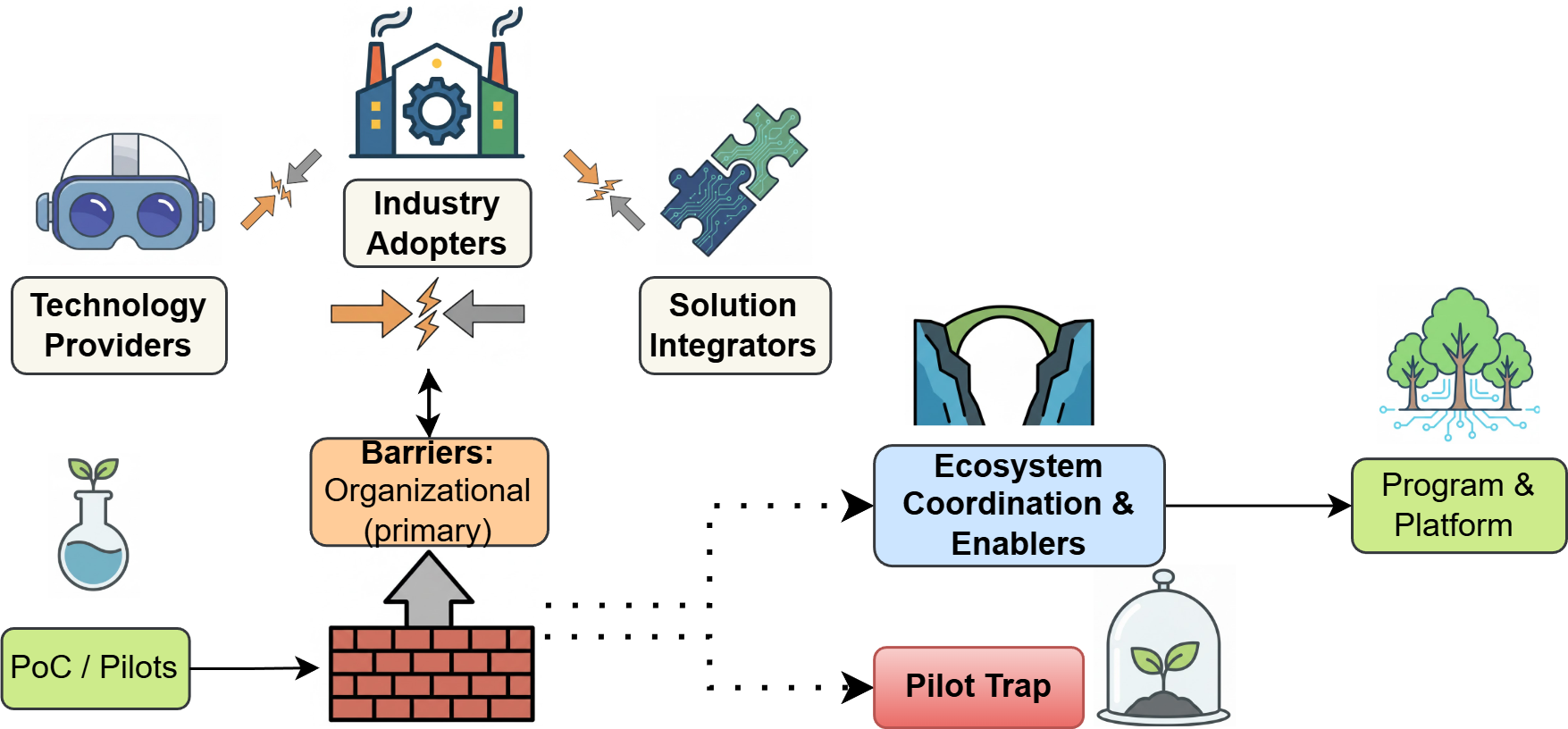} 
    \vspace{-0.3cm} 
    \caption{\textbf{Pilot Trap Mechanism.} Ecosystem misalignment creates organizational barriers (The Great Inversion). Without coordination (Bridge), pilots remain isolated (Trap) instead of scaling to platforms.}
    \label{fig:pilot_trap}
    \vspace{-0.4cm} 
\end{figure}

Extended Reality (XR) represents a pivotal shift in human-computer interaction at the intersection of digital transformation and Industry 4.0. As an immersive interface layer in digital ecosystems alongside the Internet of Things (IoT), digital twins, and artificial intelligence (AI), XR offers immense value for industrial applications, particularly in training \cite{greci_2022}, maintenance \cite{alam2025}, design \cite{harlan2020linking}, and remote collaboration \cite{Marques2022}. Market projections underscore momentum, suggesting a significant XR contribution to global GDP by 2030 \cite{PwC2019}. However, a critical paradox has emerged: despite significant investment and technical enthusiasm, enterprises struggle to transition XR from pilot projects into scaled, strategic capabilities \cite{arborxr_2025, eirmersive_2025}. Many organizations find themselves caught in a \textit{``Pilot Trap''} (e.g., Figure~\ref{fig:pilot_trap}), a cycle where technically successful proofs of concept fail to translate into operational integration or sustained value creation. These projects often remain confined to innovation labs, neither advancing to production deployment nor being formally discontinued. This suggests that the primary challenge is no longer a technological shortcoming, but a systemic adoption failure.

Existing research on XR adoption remains fragmented. While foundational frameworks such as the Technology Acceptance Model (TAM) \cite{davis1989} or Technology-Organization-Environment (TOE) \cite{tornatzky1990} offer valuable insights, they typically treat adoption as a unilateral decision made by a single firm or user, overlooking a critical reality: industrial XR adoption is an \textit{ecosystem phenomenon}. Successful integration depends on a complex interplay between three interdependent stakeholder groups: \textit{technology developers} who define capabilities, \textit{solution integrators} who translate requirements, and \textit{industry adopters} who drive organizational change. Therefore, it is essential to understand misalignments and negotiated requirements between these critical stakeholders. Current models often center on technology maturity, obscuring the coordinated nature of enterprise adoption. 

To address this research gap, we employed a qualitative, multi-stakeholder approach. Based on 17 semi-structured expert interviews triangulated across the XR value chain, we investigated the factors that influence the scaling of XR in enterprise environments. In particular, we aimed to identify the primary drivers and barriers from conflicting stakeholder perspectives, analyze the systemic friction points that lead to the Pilot Trap, and propose ecosystem-level coordination mechanisms to facilitate scaled integration. Our findings revealed a \textit{``Great Inversion''} in adoption constraints: organizational and ecosystem factors, such as change management, key performance indicator (KPI) alignment, and political resistance, have superseded technological maturity as the dominant barriers. Based on our findings, we contribute to the XR community by offering an ecosystem-centric perspective on overcoming adoption barriers in industrial settings.

\section{Related Work}
In this section, we discuss different folds of previous work on enterprise XR adoption. 

\subsection{XR in the Industrial Context}
In the context of Industry 4.0, XR has evolved from an experimental technology to a strategic interface for the smart factory \cite{kagermann2013}. Empirical studies confirm that XR acts as a catalyst for digital transformation by enabling spatial interaction with digital twins and IoT data \cite{alam2025}. The primary drivers for industrial adoption are well-documented: efficiency gains in maintenance through augmented reality (AR) overlays, reduced error rates in assembly, and significantly enhanced training transfer through virtual reality (VR) simulations \cite{fastberglund2018_xr, palmarini2018}. Particularly in occupational support, XR allows for the externalization of tacit knowledge, addressing workforce skills gaps by providing experiential learning environments that traditional methods cannot replicate \cite{hancock2020}.

\subsection{Barriers to Adoption: Technical vs. Organizational}
Early research predominantly focused on technological, financial, and ergonomic barriers, such as field-of-view limitations, battery life, costs, and cybersickness \cite{jalo2022, hussain2023}. These human factor constraints remain relevant, particularly in terms of physical and cognitive workload in dynamic workplaces.
However, recent scholarship indicates a shift. Challenges in the organizational fabric are increasingly emerging. Issues such as resistance to change, lack of IT integration, and undefined return on investment (ROI) models have emerged as critical bottlenecks \cite{badamasi2022, jalo2024}. While technical feasibility is often achieved in pilots, legacy integration in enterprise systems causes friction that purely technical studies fail to capture. 

\subsection{Current Technology Adoption Models}
To understand the technology adoption factors, previous research has traditionally relied on established frameworks. User-centric models, such as the Technology Acceptance Model (TAM) \cite{davis1989} and the Unified Theory of Acceptance and Use of Technology (UTAUT) \cite{venkatesh2003}, focus on perceived usefulness and ease of use. While extended versions have incorporated hedonic motivation and habits \cite{venkatesh2012}, they treat adoption as an isolated user decision. In contrast, firm-level frameworks, such as Technology-Organization-Environment (TOE) \cite{tornatzky1990}, analyze organizational context but often overlook inter-organizational dependencies critical to platform technologies. Recent scholarship has developed model specialization and extensions for XR technologies \cite{hussain2023, katins2024, bunz2021}, considering factors such as privacy concerns or measurement of motion sickness, indicating the need to examine specific perceptual, physiological, psychological, and contextual factors for XR technologies, which the general technology acceptance constructs can not capture alone. 

\subsection{Organizational Change Management and Adoption} 
While user-centric adoption models focus on individual acceptance, organizational adoption of transformative technologies requires coordinated change management \cite{by2005, weiner2009}. Organizational readiness to change, encompassing resource availability, leadership commitment, and workforce engagement, emerges as a critical prerequisite for sustained technology deployment \cite{weiner2009}. Recent research on organizational readiness for digital transformation emphasizes that technology procurement alone is insufficient; organizations, especially in traditional sectors,  must assess and strengthen their capacity to absorb change across individual, team, and institutional levels \cite{chanias2019}. In the XR context, user resistance, insufficient change management, and unclear value propositions are increasingly recognized as organizational barriers distinct from device-level constraints \cite{jalo2024}. This finding suggests that XR adoption models must integrate change management frameworks to explain why technically viable pilots fail to scale. The gap between individual user acceptance (e.g., captured in TAM/UTAUT) and organizational adoption (which requires sustained institutional change) remains underexplored in the XR literature. 

\subsection{The Ecosystem Gap}
A critical research gap exists in the literature regarding the \textit{ecosystem dynamics} of XR. Adoption is rarely a unilateral decision by a single firm; it is a negotiated process involving technology developers, solution integrators, and industry adopters. Current models tend to be single-actor-centric \cite{davis1989, venkatesh2003, tornatzky1990}, failing to explain systemic phenomena such as the Pilot Trap, where aligned incentives between these groups break down, and different barriers block the adoption. This study addresses this gap by triangulating perspectives across the entire value chain to reveal how ecosystem misalignments, rather than hardware maturity, act as the binding constraint on scaling, how different barriers are evolving between these group perspectives, and what benefits it can offer to understand ecosystem coordination. 

\section{Methodology}
To investigate the ecosystem dynamics of XR adoption, we employed a qualitative, exploratory research design. We conducted semi-structured expert interviews to capture convergent and divergent perspectives across the value chain. In the following sections, we provide details on our recruitment process, participants, data collection, and data analysis. 

\subsection{Recruitment, Participants, and Data Collection}
We used purposive sampling to recruit experts in senior leadership or strategic roles from different sectors. Participants were identified and contacted through professional networks (e.g., LinkedIn), direct outreach,
and through specialized events related to XR. To ensure ecosystemic validity, the sample was triangulated across three stakeholder groups: \textbf{Industry Adopters:} Organizations actively implementing XR (e.g., Automotive, Healthcare, Energy), \textbf{Technology Developers:} Hardware and software providers defining the platform capabilities, and \textbf{Solution Integrators:} Intermediaries adapting technology for enterprise needs.

To this end, we recruited 17 participants and conducted semi-structured interviews between May and October 2025, with an average duration of 47 minutes. Interviews were primarily conducted in English ($n=12$); however, a subset ($n=5$) was conducted in German upon the participants’ request to enable richer, more nuanced data capture. We refer to interview participants as P1--P17 throughout the paper to preserve confidentiality while maintaining transparency and provide detailed participant demographics in Table \ref{tab:participant_demographics}. The data collection process began with the collection of consent forms. Among our interviews, 15 of them were remote, and two were in-person. Our semi-structured interviews were guided by main questions for the main domains, starting with understanding the application contexts and expected values, then moving on to experienced barriers and whether there are any solutions against, the organizational context, and anticipations for these technologies in the ecosystem. Those were then followed up by emerging spontaneous questions based on the participants' responses. The protocol was adapted iteratively based on emerging themes during the first five interviews (pilot phase), ensuring relevance across stakeholder groups while maintaining consistency.

\begin{table*}[t]
  \caption{Participant demographics. Company names anonymized to protect participant confidentiality. F.500 = Fortune 500 company; Global = $>$5000 employees; Mid-size = 1000–4999; SME = 50–1000; Small Org = $<$50. P5 did not prefer to respond to education and age. All participants reported involvement in strategic decision-making (four indirectly, the rest directly).}
  \label{tab:participant_demographics}
  \centering
  \scriptsize 
  \begin{tabularx}{\textwidth}{l l c c l X}
    \toprule
      \textbf{ID} & \textbf{Company Type \& Scale} & \textbf{Gender} & \textbf{Age} & \textbf{Education} & \textbf{Role} \\
    \midrule
      P1 & German automotive (F.500) & M & 39 & Master's & Solution Architect for XR/Spatial Tech \\
      P2 & Global professional services (F.500) & M & 33 & Master's & Associate Manager \\
      P3 & German engineering company (F.500) & M & 35 & Master's & Product Software Developer / Lead \\
      P4 & Asian consumer electronics (Mid-size) & M & 43 & Master's & Commercial Director \\
      P5 & German automotive (F.500) & M & -- & -- & Innovation Manager: HR \\
      P6 & European software/IT company (SME) & M & 39 & Doctorate & Business / Support Unit Lead \\
      P7 & German industrial automation (F.500) & M & 49 & Master's & Domain Lead Service Lifecycle \\
      P8 & European XR/AR software (SME) & M & 32 & Bachelor's & CSO \\
      P9 & XR healthcare startup (Small Org) & M & 61 & Vocational & Co-Founder / Co-CEO \\
      P10 & American technology company (F.500) & M & 55 & Bachelor's & Global Solutions Architect \\
      P11 & European XR initiative (Small Org) & M & 43 & Master's & Managing Director \\
      P12 & German automotive (F.500) & M & 40 & Master's & Product Owner Product Design/XR \\
      P13 & Global professional services (Global) & M & 31 & Master's & Manager Immersive Technologies \\
      P14 & German industrial gases/Energy (F.500) & M & 26 & Master's & Digital Product Specialist \\
      P15 & African XR solutions (Small Org) & M & 32 & Master's & Program Director \\
      P16 & European XR Solutions (Small Org) & W & 54 & Vocational & CIO \\
      P17 & European medical technology (SME) & M & 31 & Doctorate & Lead AR/UX / Software Engineer \\
    \bottomrule
  \end{tabularx}
\end{table*}

\subsection{Data Analysis}
To analyse our data, we transcribed interviews verbatim. We translated the German interviews into English using a multi-stage validation 
protocol. First, we translated each German interview using two independent AI tools to generate alternative translations, comparing the discrepancies and selecting the most semantically equivalent version. Finally, a bilingual research team of two reviewed the translations for semantic and conceptual equivalence. For the analyses, we applied \textbf{Thematic Analysis} \cite{braunclarke2006}, chosen for its suitability in identifying, analyzing, and reporting patterns (themes) across a rich qualitative dataset, allowing insights to emerge directly from the participants’ narratives \cite{ahmed2025, braunclarke2021, clarkebraun2013}. To ensure trustworthiness and mitigate researcher bias, the entire dataset was double-coded by two researchers. First, two researchers independently coded two pilot interviews to develop and validate the coding framework. For coding units, inter-coder agreement on codes was approximately 84\%. Discrepancies were resolved through discussion and consensus, resulting in the establishment of the final thematic framework and its corresponding themes. Then, the first researcher and another researcher independently coded all 15 interviews using the established framework and themes. Using the percent average method, inter-coder agreement was 82\% at the code level and 94\% at the theme level, demonstrating robust consistency in framework application. Similarly, we resolved discrepancies through discussion and negotiated consensus. 

\section{Results}
Our analysis revealed distinct patterns in how enterprise stakeholders perceive value, navigate barriers, and manage ecosystem relationships, which we present in the following. Table~\ref{tab:participant_frequency} indicates the frequency of the categories that occurred in our interviews. 

\begin{table}[h]
\small
\centering
\caption{Participant mentions by topic category (n=17 experts). Frequency indicates the number of interview participants who mentioned each theme. The most critical insights from these topics are discussed in further detail.}
\begin{tabular}{lclc}
\hline
\textbf{Topic} & \textbf{n} & \textbf{Topic} & \textbf{n} \\
\hline
Training & 16 & Barriers: User Reservations & 8 \\
Design, Prototyping, Maintenance & 14 & Problem First Scoping & 4 \\
Efficiency and Quality & 16 & Organizational Sponsorship & 7 \\
Intangible Benefits & 9 & IT and Security & 5 \\
Barriers: Organizational & 15 & Staged Scaling & 5 \\
Barriers: Financial & 12 & User Experience Quality & 6 \\
Barriers: Technological & 7 & Ecosystem Coordination & 6 \\
Barriers: Data Privacy \& Trust & 9 & Change Management & 6 \\
\hline
\end{tabular}
\label{tab:participant_frequency}
\end{table}

\subsection{Application Contexts and Occupational Outcomes}
\label{subsec_4_1}
Enterprise XR adoption focuses on three primary industrial applications: training, design/prototyping, and maintenance support. A clear hierarchy emerges, with training as the universal anchor use case delivering immediate occupational and financial returns.

\subsubsection{Training: Hazard Mitigation and Workforce Standardization}
Training emerges as the highest-priority application across all stakeholder groups. This outcome reflects occupational transformation: XR enables safe, standardized, and scalable skill development that traditional methods cannot match. 

\textbf{Hard Skills in Hazardous Contexts:} The primary industrial use case involves technically complex or dangerous tasks where real-world training carries unacceptable safety or cost risks:

\begin{quote}
\textit{``In VR/AR training, the classic use case is dangerous situations you don't want to train in real life. Whether it's working at height or fire scenarios, that's where VR/AR makes a lot of sense.''} (P2)
\end{quote}

Industry adopters emphasize a critical occupational benefit: standardization of procedures across distributed teams. Rather than individual trainers instructing in a non-standardized way, XR delivers uniform, repeatable training:

\begin{quote}
\textit{``In the plant training environment, a lot of that training can be done digitally by interacting in a 3D VR environment. This allows everyone to complete training whenever they have time, and everyone learns the same way.''} (P1)
\end{quote}

This consistency yields downstream occupational improvements: \textit{``\textbf{better standard operating procedures (SOPs)} leading to higher product quality, and reduced process time'' (P4)}. Additionally, XR training enables \textbf{rapid upskilling} of less-experienced workers in skill-scarce environments as XR quickly trains and allows \textit{``faster memorization, and higher knowledge retention.'' (P7)}. Furthermore, it \textbf{eliminates production disruptions} through virtual training instead of on real machinery in 24/7 operations, where \textit{``even a small disruption could cause millions in losses'' (P3)}.

\textbf{Soft Skills and Behavioral Training:} Emerging applications address communication and safety-critical interpersonal skills, such as public speaking and presentation practice, highlighted by multiple experts (e.g., P4, P5, P9, P13), broadening XR's value beyond technical instruction.

\subsubsection{Design, Prototyping, and Maintenance}
The second industrial application involves design validation and digital twin creation. This domain delivers occupational impact through employee preparation for operations and accelerated time-to-market.

A critical occupational benefit emerges: employees can practice on digital twins before equipment arrives, accelerating operational readiness. Experts noted that workers ``could practice assembly, maintenance, and prepare for the whole operation'' before physical systems were deployed (P2). For capital-intensive sectors (aerospace, healthcare), digital twins eliminate costly physical training equipment, reducing costs to a fraction of traditional models (P12). 

Digital twins integrate with training, maintenance, and planning workflows, creating systemic value across the product lifecycle (P4). For long-cycle industries (e.g., seven-year vehicle development), even modest acceleration of design-to-production cycles yields substantial business impact (P12).

Hands-free maintenance support via smart glasses represents a third mature application, providing real-time guidance for complex field operations, particularly relevant for distributed occupational contexts. However, pragmatic adopters emphasize selectivity: AR complements complex tasks requiring expert judgment, not routine procedures. As one expert noted, ``AR complements specific documentation in sufficiently complex areas. You have to think carefully: where do I use AR support? Not for changing a wheel.'' (P7) This reflects a mature perspective: technology enhances but does not replace human expertise.

\subsubsection{Occupational Outcomes: Efficiency, and Quality}
Through our analyses, three primary occupational and business outcomes emerged.

\textbf{Training Efficiency and Knowledge Standardization:} 
Organizations report significant reductions in training duration and substantially higher knowledge retention compared to traditional methods (e.g., real fire extinguisher or pure theory) (P4, P5). Global standardization, achieved without instructor travel, eliminates geographic barriers to occupational consistency (P14).

\textbf{Operational Readiness and Time Compression:} 
Digital twin preparation accelerates equipment commissioning and product development cycles, with documented cases reducing six-month processes to three months and preventing costly operational errors, reducing time to market (P2, P10, P12).

\textbf{Quality Improvement, Error Prevention \& Cost Reduction:} 
Standardized VR training reduces critical errors in high-stakes occupational environments (healthcare, equipment operation, assembly), translating to reduced rework, safety incidents, and disruptions (P2, P3, P4). Many experts have mentioned cost reductions through quality improvements or by saving travel costs (e.g., P5, P6, P7, P10, P17).

\subsubsection{Sectoral Divergence and Workforce Motivation}
Critical differences emerge across sectors. Innovative sectors (e.g., automotive, manufacturing) identify multiple productive use cases spanning training, design, and operations. Conservative sectors (e.g., energy, traditional industries) remain focused on training as the primary mature application (P14), indicating that occupational readiness for XR varies by sector.

Beyond quantifiable metrics, participants identified intangible occupational value. Investing in immersive training signals organizational sophistication and employee value, supporting talent attraction and retention in skill-scarce markets (P9). Multiple adopters reported higher employee motivation when XR replaced outdated training methods (P6), suggesting that the occupational adoption of XR impacts both performance outcomes and workplace culture. 

\subsection{Barriers to Occupational XR Adoption}
\label{subsec4_2}
Enterprise XR adoption faces critical barriers that prevent scaling from isolated pilots to sustained occupational deployment. Contrary to industry narratives emphasizing hardware maturity, our analysis reveals that \textbf{organizational and financial barriers}, not technological limitations, constitute the binding constraints. This finding represents a fundamental inversion: where stakeholders expected technology to be the bottleneck, organizational readiness has become the primary blocker. 

\subsubsection{Organizational and Change Management Barriers}\label{subsubsec4_2_1}

\textbf{Awareness and Understanding Gaps:} 
The foundational barrier begins with awareness. Many organizations lack an experiential understanding of occupational impact. A technology provider emphasized that ``The first and largest barrier is awareness. Many companies are unfamiliar with VR, or if they are aware of it, they have never used a headset. They understand it on paper, but lack awareness of the impact on worker training and safety'' (P4). 

\textbf{Change Management and Organizational Readiness:} 
Solution integrators universally identify change management, not hardware, as the actual adoption bottleneck. Successful occupational adoption requires three coordinated elements: \textbf{(1)} operational champions understanding occupational benefits, \textbf{(2)} IT departments equipped to manage infrastructure, and \textbf{(3)} executive sponsorship for change processes. Most organizations achieve only one or two.

\begin{quote}
\textit{``Like any change initiative, you need to invest in people and processes, not just hardware. That's where it tends to struggle. It requires a champion high enough in the decision-making structure and real work on change management processes—things many avoid because they don't want to engage workers' councils, management, and IT.''} (P11)
\end{quote}

As another integrator noted, successful companies have both operational drivers and IT support, but ``this requires management attention and real extra motivation to start something like that'' (P8).

\textbf{Decision-Maker Resistance:} 
Subtle but powerful resistance emerges from incumbent interests. Managers with career investments in legacy systems view XR as threatening. In design-heavy sectors, as in the example of a project with a leading manufacturing company in its region, ``you're not going to tell head designers aged 50, 'we're going to use a new design tool.' Why? Because it disrupts the power dynamic you've built your career around, even if it costs a hundred times more'' (P11). Healthcare exhibits similar patterns: surgeons resist tools that deviate from methods learned during medical school (P17). This human dimension explains why well-resourced pilots often fail; technical capability matters less than the emotional investment of stakeholders in success.

\textbf{IT Infrastructure as Enabler or Blocker:} 
IT departments emerge as critical but overlooked actors. Device management, network policies, and infrastructure readiness can enable or block adoption entirely. Companies have ``different IT security concepts and different infrastructures'' (P2). Without IT support for device management and access, ``there are challenges connecting glasses to products, updates, and access'' (P5). Device management failures represent a frequently underestimated project terminator, particularly in conservative sectors like healthcare (P16).

\subsubsection{Financial and Success Measurement Barriers}
\textbf{ROI Measurement Complexity:} 
Organizations face profound challenges justifying XR investments. The barrier is not the cost itself but rather measurement ambiguity. Hard-skill training ROI is more measurable (error reduction, training time, performance gains), but soft-skill ROI (communication, safety behaviors) is strategically essential yet methodologically ambiguous. A critical geographic insight emerged: European organizations struggle with KPI definition, whereas adopters in the US establish clear before and after measurements:

\begin{quote}
\textit{``In Europe, KPIs are often not well-defined. In the US, they conduct interventions before and after and measure success explicitly. If you don't define clear KPIs upfront, after a 6–18 month pilot, you can't say whether to roll out because you haven't defined success criteria.''} (P4)
\end{quote}

Without baseline metrics, pilots generate ambiguous results, blocking the scaling decisions. 

\textbf{Scale Economics and the Missing Middle:} 
XR adoption exhibits significant scale advantages, creating economic bifurcation between large enterprises and SMEs. XR economics favor large enterprises. ``A large multi-national retailer deployed around 20,000 headsets; another major e-commerce platform has thousands, enabling long-term ROI'' (P4). Costs are amortized across thousands of users, making business cases feasible. Small- and medium-sized enterprises cannot invest at this scale and remain trapped in ``minimum viable pilot'' mode. This situation creates market bifurcation: large players adopt at scale, while SMEs remain cautious. Consequently, ``software companies can't charge €50/month because they don't have 20,000 customers. Most have far fewer, so prices remain high. Broader SME adoption would drive down software costs'' (P8), but the missing middle makes that impossible. 

\textbf{Budget Cycles and Pilot Perpetuation:} 
XR requires multi-year amortization, but pilots are framed as six-month experiments, creating a mismatch. Organizations need ``18–24 month horizons and must believe in the technology'' to justify long-term business planning (P1). However, ``when times get tough, those budgets get cut. Unless companies fully integrate XR into their processes and human resources (HR) decision-making, it will remain perpetually a pilot. Many change initiatives are incomplete—multiple pilots, but never fundamental process change'' (P11). This issue creates a structural trap: pilots become easy to cut when budgets tighten.

\subsubsection{Secondary Barriers: User Experience and Technology Stability}
While organizational and financial barriers dominate, two secondary barriers have emerged and require attention. 

\textbf{User Experience and First Impressions:} 
First-impression quality disproportionately influences adoption trajectories. Poorly designed early XR applications created enduring resistance. ``Cybersickness is 9 out of 10 times bad design'' (P6), and negative experiences create lasting impressions. However, this barrier diminishes as hardware improves and design practices mature. Critical to occupational adoption is ensuring end-user involvement and ease-of-use from pilots' outset: ``It needs to get acceptance by users. It is important to include employees who are not confident with digital technologies and make it easier, more intuitive to use'' (P3). Physical barriers (e.g., device weight, hygiene, and cosmetic concerns) are solvable through organizational commitment.

\textbf{Device Manufacturer Roadmap Uncertainty:} 
Rapid device manufacturer changes create strategic uncertainty, deterring enterprise commitment. Industry adopters hesitate to commit to platforms when manufacturers pivot rapidly or discontinue products. ``HoloLens was discontinued, divisions downsized. Other players, like Meta, focus on consumer markets and operate like startups—pivoting constantly. Industrial contexts require continuity and stability. You can't push updates that break everything; you need a 90-day grace period to test if apps function properly'' (P1). In regulated sectors like healthcare, ``What is lacking is an AR headset optimized for medical contexts. As long as there is too much change, it's unreliable technology'' (P17). This roadmap uncertainty acts as a secondary constraint on scale adoption, particularly in risk-averse sectors requiring long-term technology partnerships. 

\textbf{Data Governance and Trust:} 
Geopolitical concerns about device manufacturers (Meta, Apple, Pico) create vendor constraints in regulated markets, as the confidentiality of business content is key for industry adopters (P1, P6, P8, P10, P11). Legitimate employee privacy concerns about surveillance and eye-tracking data require transparent governance. For occupational contexts, data governance becomes a human factors issue: employees who perceive their eye movements are being tracked without consent experience reduced psychological safety and adoption resistance, directly amplifying the first-impression barriers (P6). Healthcare and energy sectors face additional compliance requirements (General Data Protection Regulation, Health Insurance Portability and Accountability Act-like frameworks) that mandate explicit data processing agreements and retention limits before pilots can begin (P16). 

However, this barrier is solvable: organizations that implement streaming architectures (processing data locally on devices), two-factor authentication, and explicit employee consent and data retention policies can effectively mitigate risks (P1, P15). The critical success factor is not technological but organizational, ensuring IT, legal, occupational safety, and worker representation teams align on governance standards and communicate these transparently to employees before pilots begin. This alignment prevents pilots from being blocked at the governance stage and supports the employee trust essential for sustained adoption. 

\subsubsection{Ecosystem Misalignment: Why Barriers Persist Unsolved} \label{subsubsec4_2_4}
The most consequential insight is that stakeholder groups perceive barriers differently, creating systemic friction that prevents the development of coordinated solutions. 

\textbf{Divergent Barrier Priorities:} 
Technology providers prioritize awareness and marketing. Adopters identify change management and organizational readiness as binding constraints. Integrators emphasize implementation quality and customization. This divergence creates a mismatch: providers optimize for technical features, while adopters need support for organizational transformation. Integrators sit between, translating but underresourced. 

A striking consensus emerges: organizational and financial barriers far exceed technical ones as adoption constraints. Yet, providers continue to invest in hardware improvements while organizational change management remains underresourced. As one provider acknowledged, ``Technical problems can be solved. I don't think there is a real blocker when it comes to technology'' (P10). This supply-demand mismatch explains why incremental hardware improvements do not accelerate adoption.

\textbf{The Pilot Trap Mechanism:} 
A substantial portion of industrial XR projects enter a perpetual ``waiting room'' state—technically viable but strategically disconnected. One adopter quantified this starkly: ``95\% of industrial XR projects end up in the waiting room. They're good projects, but companies face cultural problems. People need to believe this technology creates value`` (P14). Until that belief changes and projects connect to operational transformation, they remain experiments. 

Pilots succeed technically (employees complete training, efficiency improves) but fail strategically because they do not change underlying processes, HR systems, or career incentives. When the pilot ends, everything returns to normal. Until pilots connect to organizational strategy and implement permanent process changes, they remain isolated experiments, perpetually waiting for ``better times.''

\subsection{Enabling Sustained XR Adoption in Occupational Contexts}\label{subsec4_3}

While Section~\ref{subsec4_2} identifies barriers preventing adoption, successful organizations deploy coordinated strategies addressing organizational readiness, financial clarity, and ecosystem coordination. This section synthesizes evidence on what works across stakeholder groups.

\subsubsection{Problem-First Scoping: Occupational Outcomes over Technology}
The highest-performing XR programs invert technology-first logic, and rather than ``How do we implement XR?'', they ask ``What operational problem are we solving?'' and only then evaluate whether XR is the right tool. This approach prevents capability-driven pilots that impress technically but fail operationally. ``A lot of projects get the technology, then start looking 'what can we do with this?' whereas successful projects utilize technology specifically to address a particular problem. And then the KPIs are very well defined'' (P4). 

Problem-first approach creates three downstream benefits: financial clarity (defined ROI targets), stakeholder alignment (everyone pursues the same outcomes), and measurable gates (clear transition criteria from PoC to pilot to scale). Once the occupational problem centers are defined, organizations develop repeatable ``ROI recipes'' that simplify scaling across sites and departments. This upfront rigor prevents pilots from drifting into endless experimentation.

\subsubsection{Organizational Sponsorship and Champions}
As identified in Section~\ref{subsec4_2}, scaled adoption requires committed leadership at multiple levels. The pattern is consistent: programs fail when championed only by IT departments or innovation teams; they succeed when sponsored by business leaders with budget accountability and operational champions who have credibility within their teams. ``The best is to identify a champion in a company, who wants to achieve something'' (P9) and ``they have to be at that right level where there’s still more for them to go up, but they’re high enough where they have access to decision making and resources. Anyone out of that sweet spot is just not very useful if you want to sell a project to a company'' (P11). 

Successful adopters have champions who want to achieve something, believe in these technologies, and allocate budget, time, and resources to industrialize them across the organization. These champions drive cross-functional coordination, aligning operations, HR, IT, and production around XR integration. Without executive air cover, it is more possible that pilots remain isolated (P1, P9, P11, P12).

Technology providers note the difference: ``C-level or VP-level sponsors execute faster and scale broader. They unlock budget, align departments, and shield pilots from short-term disruptions. So this is the best way into the organization'' (P10). Operational champions, internal advocates modeling adoption, are equally critical. As adoption cannot be mandated, champions with peer credibility drive grassroots acceptance.

\subsubsection{IT and Security as Co-Designers, Not Gatekeepers}
IT departments often perceive XR as a risk or burden rather than an enabler. Successful organizations proactively engage IT and security as co-designers of governance and infrastructure. This approach prevents 'surprise compliance requests' and significantly reduces friction.

Adoption fails without proper infrastructure. As one adopter emphasized: ``When we start new XR projects, we work with IT, cybersecurity, data protection, privacy, and works council from the beginning. If you don't have IT that manages VR glasses—connecting to products, managing updates, and access—there are immediate challenges'' (P5). Another provider stressed: ``Get IT people on board at the start. There's no point kicking off a project, acquiring devices, then trying to scale without IT support'' (P10).

Clear IT and security policies upfront reduce friction, so that vendors are familiar with requirements, IT departments do not face unexpected requests, and operations understand the guardrails. As one technology provider noted, various blockers are solvable through alignment between different stakeholders: 

\begin{quote}
\textit{``If a project is shown to the IT department, the IT looks into it and sees that as a difficult topic. And then again, it creates blockers that aren’t really there if you are not aligned. If you reach out to the manufacturers for help, and if you use the right devices, many of these blockers can be immediately solved, within a couple of minutes.''} (P4)
\end{quote}

\subsubsection{Staged Scaling: From Validation to Institutionalization}
Successful adoption follows a disciplined progression with distinct objectives and gating criteria at each stage: 

\textbf{Proof-of-Concept (PoC):} This allows quick validation of the fundamental hypothesis: ``Can XR solve this problem?'' PoC bounds cohort size, duration, and scope. The goal is learning and risk mitigation, not full feature coverage. As one adopter noted: ``We assume we can save X in area A and Y in area B. Then it goes into proof of value with appropriate stakeholders selected in advance'' (P7). Another participant emphasized: ``Start small. Run PoC with 10 people, measure KPIs against previous methods, validate the value hypothesis'' (P10).

\textbf{Pilot:} The goal in piloting is to scale it to a real operational context (one site, one department) over an extended period by operationalizing governance, support, and instrumentation, and measuring KPIs together with stakeholders. As one adopter noted, ``Define KPIs with the specialist department and evaluate during development to demonstrate impact'' (P12). Including user experience metrics in addition to business metrics is also essential: ``If PoC and pilot results are strong, scaling costs become easier to justify'' (P5).

\textbf{Program \& Platform:} Once pilot metrics validate, rolling out across multiple departments and sites is necessary. ROI compounds as platform costs amortize across units, content reuses, and support becomes routine. This process involves pilots transitioning from ``interesting experiments'' to ``strategic capabilities''—and where the occupational value of training standardization, design efficiency, and maintenance support (Section~\ref{subsec_4_1}) compounds across the organization. 

\subsubsection{User Experience Quality as Adoption Lever}
Employees' first XR experience creates lasting impressions that influence their subsequent adoption for a long time. If employees' initial experience is poorly designed, such as experiencing cybersickness, using frustrating interfaces, or perceiving unclear value, they may resist adoption for years to come. In contrast, smooth, valuable first experiences create advocates. This finding implies a deliberate design of initial exposure with end-user involvement: 'Development needs to happen with the end user' (P17).

Organizations are involved in creating psychological safety (no pressure, no judgment), physical comfort (hygiene, ergonomics, dignity), and accessibility (inclusive for non-tech-savvy users). One adopter described their approach: ``We make the technology itself accessible and create an atmosphere where people can try without pressure, without performance tracking, without fear of looking ridiculous. We create a safe space to explore before discussing problem-solving. Your solution can be excellent, but if people won't put on a headset, it won't work'' (P13).

Practical comfort details, such as replaceable face masks, disinfectants, private spaces, and inclusive design choices (e.g., mirrors increasing women's participation (P1, P10)), materially affect adoption. These require organizational commitment but yield disproportionate benefits in terms of adoption gains.

\subsubsection{Ecosystem Coordination: Clarifying Roles}
XR adoption success depends critically on role clarity and deliberate coordination. When roles blur, hand-offs fail, and pilots stall, whereas when they are clear, execution accelerates. 

The stakeholder triangle functions in a way that technology providers define platforms and roadmaps, solution integrators scope problems and ensure implementation quality, and industry adopters own the requirements and provide organizational sponsorship. Each group has distinct contributions and constraints. 

A clear role definition is essential. Industry adopters should define KPIs and success metrics upfront, not providers. One integrator articulated this: ``KPIs should come from the challenges of the client, not the solution'' (P13). Solution integrators provide customization and iterative feedback, while technology providers offer stable roadmaps and enterprise-level support. 

Yet the alignment is rare. One participant stated that ``End users need to communicate demand clearly and work with providers on shared roadmaps. Many headsets are designed for consumers, not factory floors or oil rigs. That's a conversation users and providers must have'' (P8).

Sophisticated adopters establish coordination mechanisms. Another participant noted: ``We have tech offices at critical locations, working directly with device makers and software providers. The aim is to adapt technologies quickly, build local ties, and react rapidly to market changes. We work with global technology providers on new use cases and glass optimization for enterprise deployment'' (P12).

\subsubsection{Change Management: The Primary Adoption Lever}
Across all stakeholders, organizational change emerged as the primary adoption lever, yet it is systematically underresourced. Change management should constitute a larger program investment, not an afterthought.

This direction includes communications planning, manager enablement (i.e., modeling adoption), peer learning cohorts, resistance management, and incentive alignment. Without these, even excellent XR systems are adopted sporadically and eventually abandoned. This direction was stated as ``It's a change management process—you need to motivate people, maybe financially, to spend time on this. If not addressed, technology champions face tremendous frustration and hurdles'' (P8).

As one expert (P11) emphasized, this misalignment highlights that the hardware is an easy investment, but it struggles with investing in people and processes (i.e., in Section~\ref{subsubsec4_2_1}). Identifying change management as a core workstream and funding it appropriately is essential for innovation projects to succeed.

\section{Discussion}
This study reveals a fundamental reevaluation of enterprise XR adoption constraints and identifies systemic barriers to occupational deployment that existing technology adoption models often overlook. The findings, drawn from 17 multi-stakeholder experts, challenge technology-deterministic narratives and expose critical gaps between what the XR industry is optimizing for and what organizations need to sustain occupational integration.

\subsection{The Great Inversion: Organizational Readiness as the Binding Constraint}
A striking finding emerges across all three stakeholder groups: technology barriers no longer constitute the primary bottleneck to adoption. Instead, organizational readiness, encompassing change management, executive sponsorship, IT alignment, and financial governance, has become the binding constraint.

This inversion directly challenges foundational adoption frameworks. The TOE model treats Technology, Organization, and Environment as coequal contexts. Our findings suggest that for mature yet emerging technologies like XR, this weighting may be misleading. Hardware and software have reached a level of maturity that is ``good enough'' for most occupational use cases. The primary variability now lies in organizational factors: whether adopters have champions with budget authority, whether IT co-designs infrastructure, whether KPIs are defined upfront, and whether change management is resourced as a core workstream.

Critically, this inversion is not universal. Automotive organizations, within structured digital transformation programs, show high organizational readiness. Healthcare organizations, constrained by regulatory conservatism and fragmented decision-making processes, exhibit lower readiness. This sectoral variation suggests that occupational XR adoption is contingent, not universal, a finding that extends adoption theory toward contingency-based models accounting for institutional logic beyond technology maturity. 

\subsection{The Pilot Trap: An Ecosystem-Level Failure}
A critical pattern emerges from our analysis. As an expert stated based on his experience, ``95\% of projects are going to the waiting room'' (Section \ref{subsubsec4_2_4}), and while the number is one expert’s characterization and not an empirical claim, the phenomenon itself (technical success does not translate into strategic scaling) is well-supported and highlighted multiple times. This finding reveals a critical failure in established adoption theory. Pilots succeed technically, employees complete training, efficiency improves, but the process fails strategically, as pilots do not transition to sustained operational deployment.

Existing adoption models (e.g., TAM, UTAUT, TOE) cannot explain this effect because they are single-actor or single-firm centric. They address individual acceptance or organizational context, but not the ecosystem-level failure that emerges when multiple interdependent actors optimize locally without systemic coordination.

We identify the root cause as a misalignment between stakeholder priorities and incentives. Technology providers optimize for technical feasibility and platform standardization. Industry adopters struggle with organizational readiness, requiring customization and change management support that providers do not offer. Solution integrators occupy the middle, translating between worlds but underresourced for the transformation work needed. The outcome is predictable: a technically successful pilot is achieved, but the adopter organization lacks the necessary processes, including redesign, human resources integration, capital budget alignment, and effective change communication, to scale its operation. Providers interpret this as ``adopter underperformance''; adopters experience it as ``we need organizational transformation, not just technology.'' This disconnect is the Pilot Trap.

The mechanism is addressable with a deliberate ecosystem coordination from project inception. When different roles and demands are well communicated, meaning that adopters define KPIs and success metrics upfront, integrators explicitly manage organizational change and measurement discipline, and providers commit to enterprise stability and support, it enables stakeholder alignment. As a result, pilots transition to programs and then to platforms. In contrast, when coordination is absent, pilots remain as isolated experiments, perpetually waiting for ``better times'' or budget cycles.

\subsection{Implications for Occupational XR in Industrial and Workplace Contexts}
Our findings reframe occupational XR adoption as fundamentally an \textbf{organizational and human factors challenge}, not a technology challenge. This framing has direct implications for how the community approaches research and practice, which are as follows. 

\textbf{Scope Expansion for Ergonomics and Human Factors:} 
Traditional occupational ergonomics addresses device-level factors, such as field of view, weight, and cybersickness. Our study reveals that organizational factors equally shape employee experience, influencing whether they feel valued (through champion modeling), perceive control (through clear KPIs and success communication), trust the system (through IT governance and privacy transparency), and whether their first impressions are positive (through safe, non-judgmental onboarding). Occupational XR adoption researchers should integrate organizational ergonomics, including policies, roles, and incentives, alongside device-level ergonomics. 

\textbf{Occupational Training Standardization as a Strategy:} 
The universal anchor use case is training. Its value is not inherent to XR but arises from standardized procedures, repeatable curricula, and measurable competency. Organizations that institutionalize training as a core occupational process, with KPIs, governance, and executive support, scale adoption regarding the problem-first approach. In contrast, those who treat it as a technology pilot tend to underperform. This finding suggests that future research on XR-based competency systems should focus on designing training programs that integrate with occupational qualification frameworks, safety protocols, and workforce development strategies. 

\textbf{Importance of First Impressions as Occupational Psychology:} 
Whether employees adopt XR depends disproportionately on their initial experience with it. Psychological safety (absence of performance pressure and judgment), physical comfort (hygiene, accessibility, and dignity), and perceived control (clear value and absence of surveillance) shape lasting attitudes. These factors are not associated with device design; they are rather related to organizational design. Future occupational XR research should investigate how room setup, manager enablement, and inclusive onboarding protocols affect employee adoption trajectories.

\subsection{Anticipated Evolution: AI-XR Convergence and Persistent Organizational Constraints}

Hardware evolution is moving toward lighter, more ergonomic all-day wearables (P10, P13). AI integration is becoming essential, with natural language interfaces, predictive training, and autonomous maintenance guidance (P7, P8, P15) being key examples. As people become accustomed to the ease of use of AI, integrated AI systems will become a significant factor in enhancing the usability of XR technologies for employees. Some participants also noted the advantages of having an ecosystem for technology developers, its benefits for adopters, and the importance of standards for interoperability, as seen in examples such as Android/Google or Apple (P5, P8, P12). These trajectories will significantly enhance occupational value. However, they do not remove the organizational constraints identified in this study; they amplify them.

AI-XR systems will create new governance requirements, such as data ownership, algorithmic transparency, and human-AI teaming protocols. Organizations currently struggling with XR organizational readiness will face similar challenges with AI-XR integration in the near future. The binding constraint will remain organizational, not technological. Healthcare may continue to lag due to regulatory complexity, while manufacturing may accelerate due to efficiency imperatives. The sectoral variation we observe nowadays is likely to persist, shaped by institutional logic and organizational readiness, rather than technology maturity.

\subsection{Practical Implications}
We outline several implications for three stakeholder groups and researchers below.

For \textbf{industry adopters}, it is key to treat XR adoption as an occupational transformation program, not an isolated innovation initiative. They should secure executive sponsorship tied to occupational outcomes (training standardization, employee safety, cycle time), engage IT as a co-designer from inception, invest a significant amount of XR budgets in change management, not devices. Furthermore, they should define occupational KPIs before pilots begin and establish ecosystem partnerships with integrators and providers early, building shared roadmaps and clear role definitions to ensure seamless integration. 

\textbf{Solution integrators} should focus on differentiating between organizational transformation and technology implementation. They should offer playbooks for problem-first scoping, KPI measurement, and staged scaling (PoC → Pilot → Program → Platform). They should explicitly include resource change management and stakeholder alignment, and position themselves as 'transformation partners' bridging technology providers and organizational reality.

For \textbf{technology providers}, analysis shows that stability and enterprise support now drive competitive advantage more than features. Publishing multi-year hardware roadmaps with update grace periods in communication with other stakeholders, while providing data governance and privacy reference architectures, will ensure easier adoption. Investing in customer success teams that support organizational readiness, not just technical implementation, is also identified as a key factor.

\textbf{Human factors and ergonomics researchers} should expand occupational research beyond device-level factors. Investigating organizational policies, champion effects, first-impression design, and governance as human factors, while conducting comparative studies across sectors, can provide hidden insights into how institutional logic shapes employee adoption. Exploring XR-specific change management frameworks informed by occupational psychology and implementation science is also essential. 

\subsection{Limitations and Future Research}
While our study provides in-depth insights into adoption mechanisms, it has a few limitations. Firstly, the sample is primarily European and weighted toward advanced adopters. Furthermore, while we have some sectoral variation (automotive vs. healthcare in maturity), not all the sectors are well-represented. 

Considering our limitations, future research is needed. Future studies should focus on \textbf{(i)} the quantitative validation of the great inversion with longitudinal, multi-site, quantitative studies measuring organizational versus technological constraints and their relative impact on adoption success over time. Additionally, research is needed on \textbf{(ii)} the comparative case studies of pilot escape by identifying specific governance structures, change management interventions, and ecosystem coordination mechanisms that enable organizations to transition from pilot to sustained deployment. Furthermore, \textbf{(iii)} occupational XR design research with controlled studies on first-impression design (safe spaces, onboarding formats, inclusive accessibility) and their long-term effects on employee adoption, complementing traditional ergonomics and cybersickness research, is needed. Lastly, \textbf{(iv)} cross-industry comparative research (e.g., healthcare, manufacturing, energy, construction), identifying how regulatory environments, safety cultures, and occupational structures shape XR governance and adoption, is also essential. 

\section{Conclusion}
In this work, we explored organizational readiness and ecosystem coordination for industrial XR with a multi-stakeholder study that included 17 expert interviews. Our results reveal that occupational XR adoption is fundamentally an organizational and ecosystem challenge, rather than a technological one. Through our analysis, we identified the Pilot Trap, an ecosystem-level failure where technically successful pilots fail to scale due to stakeholder misalignment, undefined KPIs, and insufficient organizational change management, as the primary obstacle to sustained deployment. Our findings demonstrate that organizational readiness, change management, and ecosystem coordination play a more significant role in adoption success than technological maturity. We argue that successful occupational XR requires simultaneous orchestration of problem-first scoping, executive championing, IT co-design, staged scaling, user-centric design, and sustained change management in the ecosystem. By reframing occupational XR adoption as an organizational transformation process rather than technology procurement, practitioners and researchers can accelerate the transition from perpetual pilots to sustained platforms. 

\bibliographystyle{abbrv-doi}

\bibliography{template}
\end{document}